\documentclass[namedreferences]{SolarPhysics}
\usepackage[optionalrh,solaenum,natbib]{spr-sola-addons} 
\usepackage{graphicx}                    
\usepackage{amssymb}                     
\usepackage{color}                       
\usepackage{url}                         

\newcommand{\aap}{{\it Astron. Astrophys.}}

\newcommand{\apj}{{\it Astrophys. J.}}
\newcommand{\apjl}{{\it Astrophys. J. Lett.}}

\newcommand{\jgr}{{\it J. Geophys. Res.}}

\newcommand{\mnras}{{\it Mon. Not. Roy. Astron. Soc.}}

\newcommand{\solphys}{{\it Solar Phys.}}

\newcommand{\eg}{{\textit{e.g.}}}
\newcommand{\ie}{{\textit{i.e.} }}
\newcommand{\prl}{{\it Phys. Rev. Lett.}}      
\newcommand{\an}{{\it Astronom. Nachr.}}
\newcommand{\na}{{\it New Astronomy}}

\begin{document}
\begin{article}
\begin{opening}
\title{
Initiation of Coronal Mass Ejections by Sunspot Rotation
}
%
\runningauthor{T. T\"or\"ok \textit{et al.}}
\runningtitle{CME Initiation By Sunspot Rotation}

\author{
{T. T\"or\"ok}$^{1}$ \sep
{M. Temmer}$^{2,3}$   \sep
{G. Valori}$^{4}$   \sep
{A.~M. Veronig}$^{2}$   \sep
{L. van Driel-Gesztelyi}$^{4,5,6}$   \sep
{B. Vr\v snak}$^{7}$   \sep
}
\institute{
$^{1}$ Predictive Science, Inc., 9990 Mesa Rim Rd., Suite 170, San Diego, CA 92121, USA; email: {\sf tibor@predsci.com}\\
$^{2}$ IGAM/Kanzelh\"ohe Observatory, Institute of Physics, Universit\"at Graz, Universit\"atsplatz 5, A-8010 Graz, Austria\\
$^{3}$ Space Research Institute, Austrian Academy of Sciences, Schmiedlstrasse 6, A-8042 Graz, Austria\\
$^{4}$ LESIA, Observatoire de Paris, CNRS, UPMC, Universit\'e Paris Diderot, 5 place Jules Janssen, 92190 Meudon, France \\
$^{5}$ University College London, Mullard Space Science Laboratory, Holmbury St. Mary, Dorking, Surrey RH5 6NT, UK\\
$^{6}$ Konkoly Observatory, Hungarian Academy of Sciences, Budapest, Hungary\\
$^{7}$ Hvar Observatory, Faculty of Geodesy, University of Zagreb, Ka\v ci\'ceva 26, HR-10000 Zagreb, Croatia\\
}
\date{Received ...; accepted ...}
\begin{abstract}
We study a filament eruption, two-ribbon flare, and coronal mass ejection (CME) that occurred in Active Region NOAA 10898 on 6 July 2006. The filament was located South of a strong sunspot that dominated the region. In the evolution leading up to the eruption, and for some time after it, a counter-clockwise rotation of the sunspot of about 30 degrees was observed. We suggest that the rotation triggered the eruption by progressively expanding the magnetic field above the filament. To test this scenario, we study the effect of twisting the initially potential field overlying a pre-existing flux-rope, using three-dimensional zero--$\beta$ MHD simulations. We first consider a relatively simple and symmetric system, and then study a more complex and asymmetric magnetic configuration, whose photospheric flux distribution and coronal structure are guided by the observations and a potential field extrapolation. In both cases, we find that the twisting leads to the expansion of the overlying field. As a consequence of the progressively reduced magnetic tension, the flux-rope quasi-statically adapts to the changed environmental field, rising slowly. Once the tension is sufficiently reduced, a distinct second phase of evolution occurs where the flux-rope enters an unstable regime characterized by a strong acceleration. Our simulations thus suggest a new mechanism for the triggering of eruptions in the vicinity of rotating sunspots.
\end{abstract}
\keywords
{Magnetic fields, Corona -- Active Regions, Models -- Coronal Mass Ejections, Initiation and Propagation -- Sunspots, Velocity}
\end{opening}

\section {Introduction}
\label{sec:int}

Filament (or prominence) eruptions, flares, and coronal mass ejections (CMEs) are the three large-scale eruptive events on the Sun. It has become clear in recent years that they are not independent phenomena, but different observational manifestations of a more general process, namely the sudden and violent disruption and dynamic reconfiguration of a localized volume of the coronal magnetic field \citep[\eg][]{forbes00}. Whether or not all three phenomena occur together appears to depend mainly on the properties of the pre-eruptive configuration. For example, CMEs can occur without a filament eruption (if no filament has formed in the source region of the erupting flux prior to its eruption) and without significant flaring (if the magnetic field in the source region is too weak; \eg \, \citealt{zirin88}) or, in extreme cases, even without any low-coronal or chromospheric signature \citep{robbrecht09}. On the other hand, both flares and filament eruptions are not always accompanied by a CME (if, for instance, the magnetic field above the source region is too strong; see, {\eg}, \citealt{moore01}, \citealt{nindos04}, \citealt{torok05}). In large events such as the one studied in this article, however, all three phenomena are observed almost always. Such events typically start with the slow rise of a filament and/or overlying loops \citep[\eg][]{maricic04,schrijver08a,maricic09}, which is often accompanied by weak pre-flare signatures in EUV or X-rays \citep[\eg][]{maricic04,chifor07}. The slow rise is followed by a rapid acceleration and a huge expansion of the eruptive structure, which is then observed as a CME. The rapid acceleration has been found in most cases to be very closely correlated with the flare impulsive phase \citep[\eg][]{kahler88,zhang.j01,maricic07,temmer08}.

Although it is now widely accepted that solar eruptions are magnetically driven, the detailed physical mechanisms that initiate and drive eruptions are still controversial. Accordingly, a large number of theoretical models have been proposed in the past decades \citep[for a recent review see, \eg,][]{forbes10}. Virtually all of these models consider as pre-eruptive configuration a sheared or twisted core field low in the corona, which stores the free magnetic energy required for eruption and is stabilized by the ambient coronal field. The choice of such a configuration is supported by observations of active regions, which often display sheared structures (filaments and soft X-ray sigmoids) surrounded by less sheared, tall loops. An eruption is triggered if the force balance between the core field and the ambient field is destroyed, either by increasing the shear or twist in the core field or by weakening the stabilizing restoring force of the ambient field \citep[see, \eg,][]{aulanier10}.

One of the many mechanisms that has been suggested to trigger eruptions is the rotation of sunspots. The idea was put forward by \cite{stenflo69}, who showed that the order of magnitude of the energy deposition into coronal structures by sunspot rotations is sufficient to produce flaring activity \citep[see also][]{kazachenko09}.

Sunspot rotations have been known for a long time -- the first evidence, based on spectral observations, was presented one century ago by \cite{evershed1910} -- and since then they they have been the subject of numerous analyzes. Still, measurements of sunspot rotation are not straightforward, and, depending on the employed method, can give quite different results \citep[see, \eg,][]{min09}. Meticulous case studies \citep[\eg][]{zhang.j07a,min09,yan.xl09}, as well as detailed statistical analyzes \citep[\eg][]{brown03,yan.xl07,zhang.y08,li.l09,suryanarayana10} showed that sunspots can rotate significantly, up to several hundreds of degrees over a period of a few days. Interestingly, sunspots do not necessarily rotate as a rigid body, \cite{brown03} and \cite{yan.xl07} showed that the rotation rate often changes with the distance from the sunspot centre. The rotation of sunspots is commonly interpreted as an observational signature of the emergence of a flux-rope through the photosphere \citep[\eg][]{gibson04} or, more generally, as the transport of helicity from the convection zone into the corona \citep[see, \eg,][]{longcope00,tian06,tian08,fan09}. On the other hand, observations of strong sunspot rotation without signs of significant flux emergence have been reported \citep[\eg][and references therein]{tian06}, suggesting that intrinsic sunspot rotation of sub-photospheric origin exists. In such cases the rotation rate tends to be smaller than for sunspot rotations associated with flux emergence \citep[\eg][]{zhu12}. 

A number of studies have shown a direct cause--consequence relationship between higher-than-average sunspot rotation and enhanced eruptive activity. For example, \cite{brown03}, \cite{hiremath03}, \cite{hiremath06}, \cite{tian06}, \cite{yan.xl07}, \cite{zhang.y08}, \cite{li.l09}, \cite{yan.xl09,yan.xl12} and \cite{suryanarayana10} reported an apparent connection between rotating sunspots (with total rotation angles of up to 200$^{\circ}$ and more) and eruptive events. In particular, \cite{yan.xl07} attributed eruptive activity in an active region to different rotation speeds in different parts of a sunspot, whereas \cite{yan.xl08} found indications that active regions with sunspots rotating opposite to the differential rotation shear are characterized by high X-class-flare productivity. \cite{romano05} reported a filament eruption that was apparently triggered by photospheric vortex motions at both footpoints of the filament, without any sign of significant flux emergence.

Besides purely observational studies of the relationship between sunspot rotation and eruptive activity, some authors presented a combination of observations and modeling. For example, \cite{regnier06} utilized multi-wavelength observations and modeling of the coronal magnetic field of the highly flare-productive active region NOAA AR8210 to show that slow sunspot rotations enabled flaring, whereas fast motions associated with emerging flux did not result in any detectable flaring activity. Moreover, they also showed that the deposition of magnetic energy by photospheric motions is correlated with the energy storage in the corona, which is then released by flaring. Similarly, \cite{kazachenko09} analyzed detailed observations of an M8 flare--CME event and the associated rotating sunspot, and combined them in a minimum--current--corona model. They found that the observed rotation of 34$^{\circ}$ over 40 hours led to a triplication of the energy content and flux-rope self--helicity, sufficient to power the M8 flare.

Numerical MHD investigations of the relationship between sunspot rotation and eruptive activity started with \cite{barnes72}, who modeled the coronal magnetic field of a rotating sunspot surrounded by a region of opposite polarity. They found that the rotation causes an inflation of the magnetic field, and that its energy increases with the rotation angle until, when the rotation angle exceeds $\approx 180^{\circ}$, it becomes larger than that of the open-field configuration with the same boundary conditions, presumably leading to an eruption. 

MHD simulations of the formation and evolution of flux-ropes by twisting line--tied potential fields have been widely performed since then. Calculations were done by either twisting uniform fields in straight, cylindrically symmetric configurations \citep[\eg][]{mikic90,galsgaard97,gerrard02,gerrard03a} or by twisting bipolar potential fields, the latter yielding arched flux-ropes anchored at both ends in the same plane \citep[\eg][]{amari99a,gerrard04}. Most of these simulations focused on the helical kink instability and its possible role in producing compact flares and confined eruptions. \cite{klimchuk00} studied the twisting of a bipole with emphasis on the apparently uniform cross-section of coronal loops. Very recently, \cite{santos11} simulated the energy storage for the active region that was studied earlier by \cite{regnier06}. They imposed photospheric flows on an extrapolated potential field and found the formation of pronounced electric currents at the locations of the observed flare sites. The authors concluded that the main flare activity in the active region was caused by the slow rotation of the sunspot that dominated the region. 

However, none of the above studies were directly related to CMEs. \cite{amari96} were the first to show that the formation and continuous twisting of an arched flux-rope in a bipolar potential field can lead to a strong dynamic expansion of the rope, resembling what is observed in CMEs. Later on, \cite{torok03} and \cite{aulanier05} extended this work by studying in detail the stability properties and dynamic evolution of such a system. The underlying idea of these simulations is that slow photospheric vortex motions can twist the core magnetic field in an active region up to the point where equilibrium cannot be longer maintained, and the twisted core field, \ie a flux-rope, erupts (for the role of increasing twist in triggering a flux-rope eruption see also \citealt{chen.j89,vrsnak90,fan03,isenberg07}). What has not been studied yet is whether a twisting of the field {\em overlying an existing flux-rope} can lead to the eruption of the rope.

In this article, we present observations of a large solar eruption which took place in the vicinity of a rotating sunspot. We suggest that the continuous rotation of the spot triggered the eruption by successively weakening the stabilizing coronal field until the low-lying core field erupted. We support our suggestion by MHD simulations that qualitatively model this scenario.

The remaining part of this article is organized as follows. In Section~\ref{sec:obs} we describe the observations, focusing on the initial evolution of the eruption and on the rotation of the sunspot. In Section~\ref{sec:num} we describe the numerical simulations, the results of which are presented in Section~\ref{sec:res}. We finally discuss our results in Section~\ref{sec:dis}.

\section{Observations}
\label{sec:obs}

\begin{figure}[t]
\centering
\includegraphics[width=0.7\linewidth]{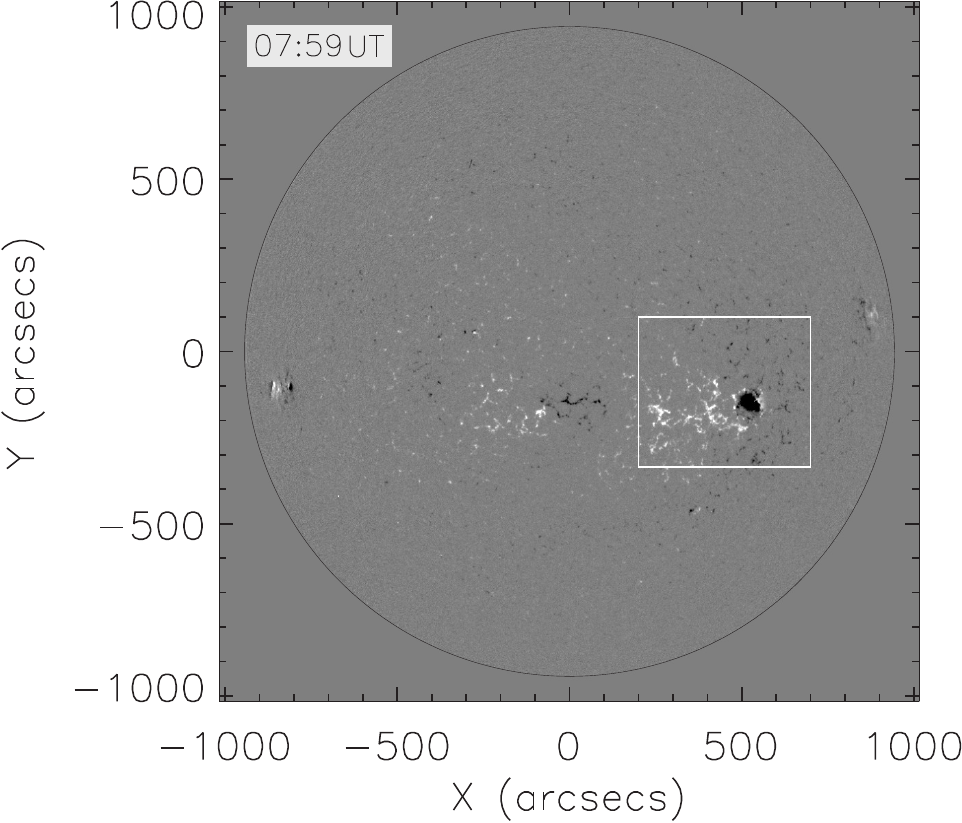}
\caption
{
Full-disk line-of-sight SOHO/MDI magnetogram recorded on 6 July 
2006, 07:59 UT. The active region under study is marked by the 
white box. 
}
\label{fig:mdi}
\end{figure}

The eruption on 6 July 2006 in active region NOAA  10898 was a textbook two-ribbon flare accompanied by a filament eruption and a halo CME, the latter being most prominent in the southwest quadrant and reaching a linear plane-of-sky velocity of $\approx 900$ km\,s$^{-1}$ \citep{temmer08}. The event was associated with an EIT wave, a type II burst, and very distinct coronal dimming regions. The flare was of class M2.5/2N, located at the heliographic position S9$^{\circ}$, W34$^{\circ}$. It was observed in soft X-rays (SXR) by GOES (peak time at $\approx$ 08:37~UT) as well as in hard-X rays (HXR) with RHESSI, with the two highest peaks of nonthermal HXR emission occurring during 08:20\,--\,08:24~UT.

The evolution of the active region in the days preceding the eruption, and in particular the rotation of the leading sunspot, can be studied using its photospheric signatures. Photospheric line-of-sight magnetograms of the region were obtained by the MDI instrument \citep{scherrer95} onboard the {\em Solar and Heliospheric Observatory} (SOHO). The active region was a bipolar region of Hale type $\beta$, consisting of a compact negative polarity (the sunspot) that was surrounded by a dispersed positive polarity, most of which was extending eastwards (see Figure~\ref{fig:mdi}). The maximum of the magnetic-field flux density in the sunspot was about nine times larger than in the dispersed positive polarity. The two polarities were surrounded by a large, ``inverse C-shaped'' area of dispersed negative flux to the west of the region. 

\begin{figure}
\centering
\includegraphics[width=1.0\linewidth]{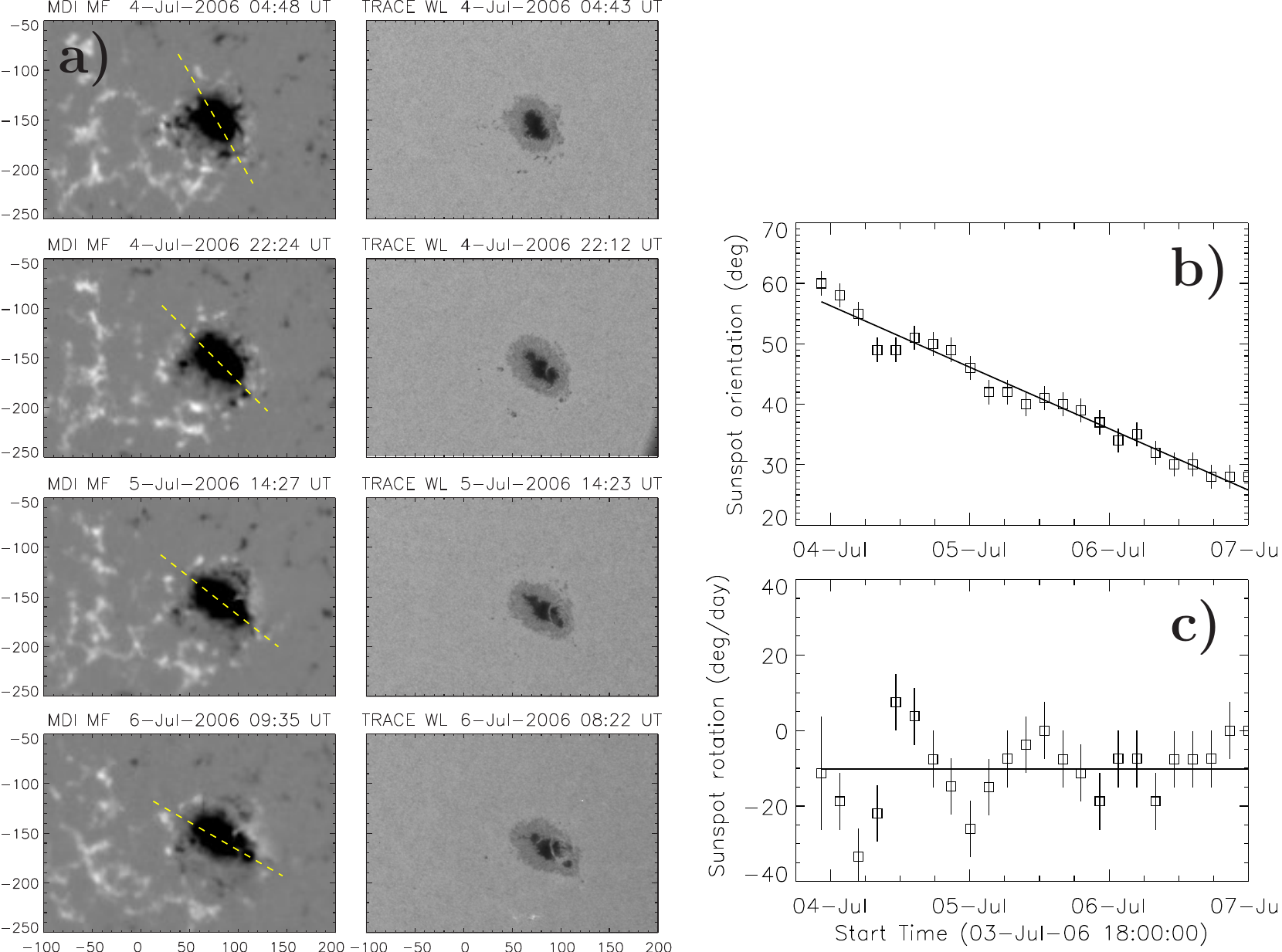}
\caption
{
(a) Representative images of the sunspot evolution during 4\,--\,6 July 2006: 
MDI longitudinal magnetic-field maps (left column); TRACE white-light images 
(right column). The TRACE image in the bottom panels corresponds to the time 
of the M2 flare (starting in soft x-rays at 8:20~UT). The dashed yellow line 
outlines the major axis of the sunspot that was used to measure the sunspot 
rotation. The corresponding SOHO/MDI movie is available in the electronic 
version of the article.
(b) Sunspot rotation determined from the MDI magnetic-field maps over the period 
3 July 2006, 22:00~UT, to 7 July 2006, 8:00~UT, showing the orientation of the 
sunspot's major axis, measured clockwise from solar East.
(c) Sunspot rotation rate in degrees per day.
}
\label{fig:spot_evolution}
\end{figure}

\begin{figure}
\centering
\includegraphics[width=0.85\linewidth]{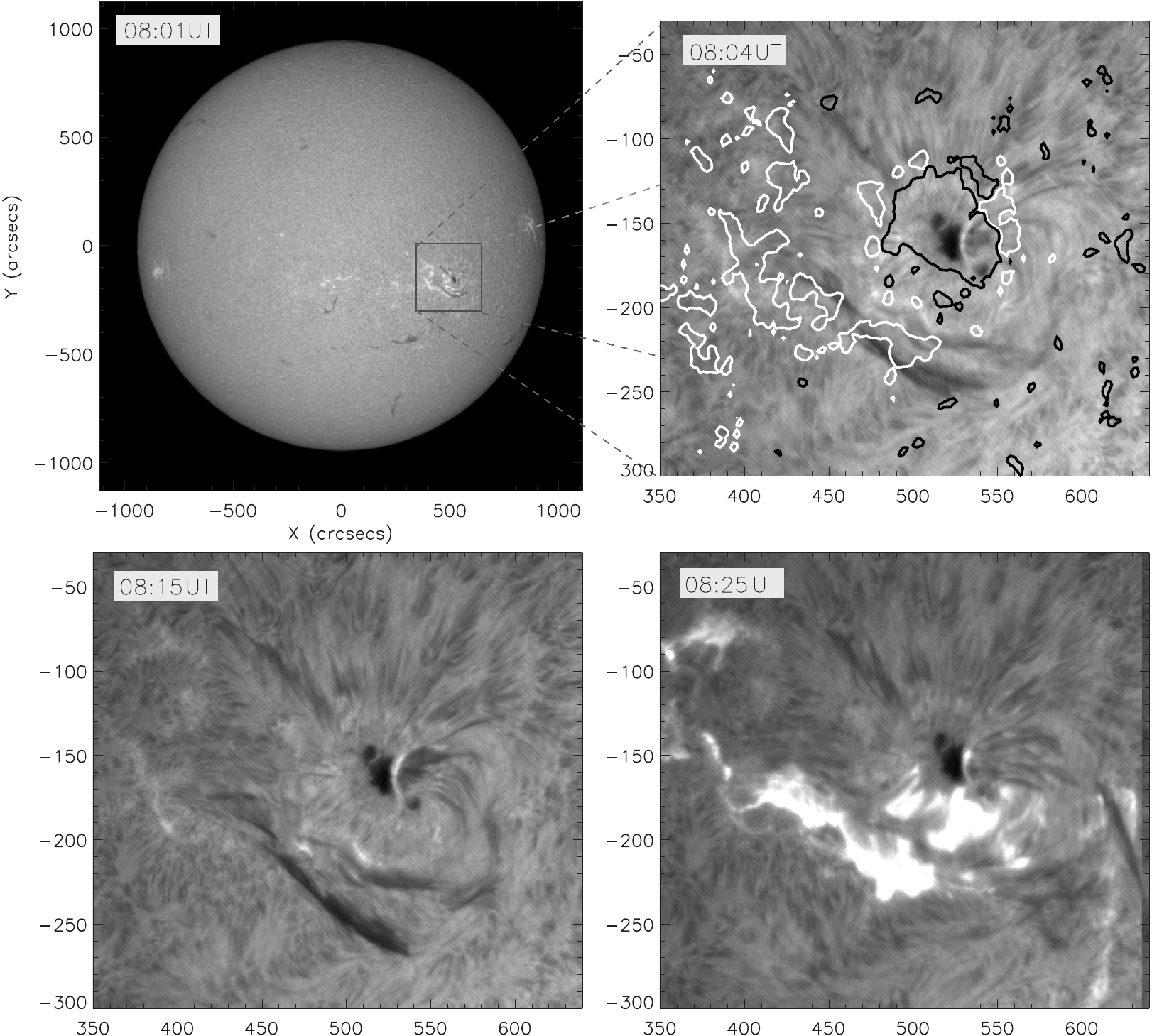}
\caption
{
H$\alpha$ filtergram sequence observed before and during the flare on 6 July 2006 by the 
Kanzelh\"ohe (full disk) and Hvar (active region area) Observatories. An apparent double 
structure of the filament is visible South of the sunspot. Contour levels of 100 G from 
an MDI magnetogram taken at 07:59 UT are added in the top right panel, with white (black) 
lines corresponding to positive (negative) values.
}
\label{fig:ha_sequ}
\end{figure}

We measured the magnetic flux of the concentrated leading (negative) and dispersed following 
(positive) polarities using a (re-calibrated) SOHO/MDI synoptic map, which preserves the resolution 
of the original observation. The map includes magnetic features close to the time of their central 
meridian passage, when projection effects of the line-of-sight magnetic fields are at minimum. The 
total magnetic flux (half of the total unsigned flux) was found to be $(2.1 \pm 0.2) \times 10^{22}$ 
Mx, with the two polarities nearly balanced [$(2.0 \pm 0.2)\times 10^{22}$ and $(-2.2 \pm 0.2) 
\times 10^{22}$ Mx for the positive and negative flux, respectively]. The error estimates reflect 
the uncertainty in determining how much of the dispersed positive and negative polarities belonged 
to the active region. The leading spot, including the penumbral area, had a mean magnetic-field 
strength (magnetic-flux density over 2340 pixels) of  390 G, reaching 1820 G when a smaller, purely 
umbral, area was considered (240 pixels). However, since the MDI response becomes non-linear in 
such a strong, and therefore dark, umbra, the core field strength there was probably higher 
($\geq$ 2000 G) \citep[see, \eg,][]{green03}. The positive dispersed plage had a much lower mean 
magnetic-field strength of about $50 \pm 10$ G, depending on the extent of dispersed positive field
measured (magnetic-flux density over 13\,060\,--\,24\,600 pixels). Positive flux concentrations (measured 
over 600 pixels) within the plage had a characteristic field strength of $220 \pm 20$ G. In summary, 
magnetic-flux measurements indicate a mere 5\% negative surplus flux in this major bipolar active 
region of $2.1 \times 10^{22}$ Mx total flux and maximum-field strengths (negative:positive) 
in a roughly 10:1 ratio.

In Figure~\ref{fig:spot_evolution}(a) we show snapshots of the sunspot evolution as observed by MDI and the {\em Transition Region and Coronal Explorer} \citep[TRACE:][]{handy99}, ranging from two days before the eruption to one day after it. The images are all differentially rotated to the first image of the series, when the sunspot was closer to disk centre. The sequence shows that the sunspot is rotating counter-clockwise during the considered period (see the Electronic Supplementary Material). From the evolution of the MDI magnetic-field maps, we geometrically determined the major axis of the sunspot and followed its evolution in time. In Figure~\ref{fig:spot_evolution}(b) we plot the sunspot's rotation angle over the period 3 July 2006, 22:00 UT, to 7 July 2006, 8:00 UT. The total rotation observed over these three days is about 30$^\circ$. The sunspot's rotation rate, determined as the temporal derivative of the rotation  measurements, yields a mean value of about 10$^\circ$ day$^{-1}$ during the considered time span (Figure~\ref{fig:spot_evolution}(c)). For comparison, we determined the rotation also from the TRACE white-light images and found no significant differences.

The flare and the filament eruption were observed in full-disk H$\alpha$ filtergrams by the Kanzelh\"ohe Observatory and, over a smaller field-of-view around the active region, by the Hvar Observatory (Figure~\ref{fig:ha_sequ}). These observations reveal that the filament consisted of a double structure before and during the eruption \citep[for a similar case of such a double-structured filament, see][]{liu.r12}. Significant rising motions of the filament could be seen from about 08:23 UT on. The H$\alpha$ flare started by the appearance of very weak double-footpoint brightening at 08:15 UT.

\begin{figure}
\centering
\includegraphics[width=1.\linewidth]{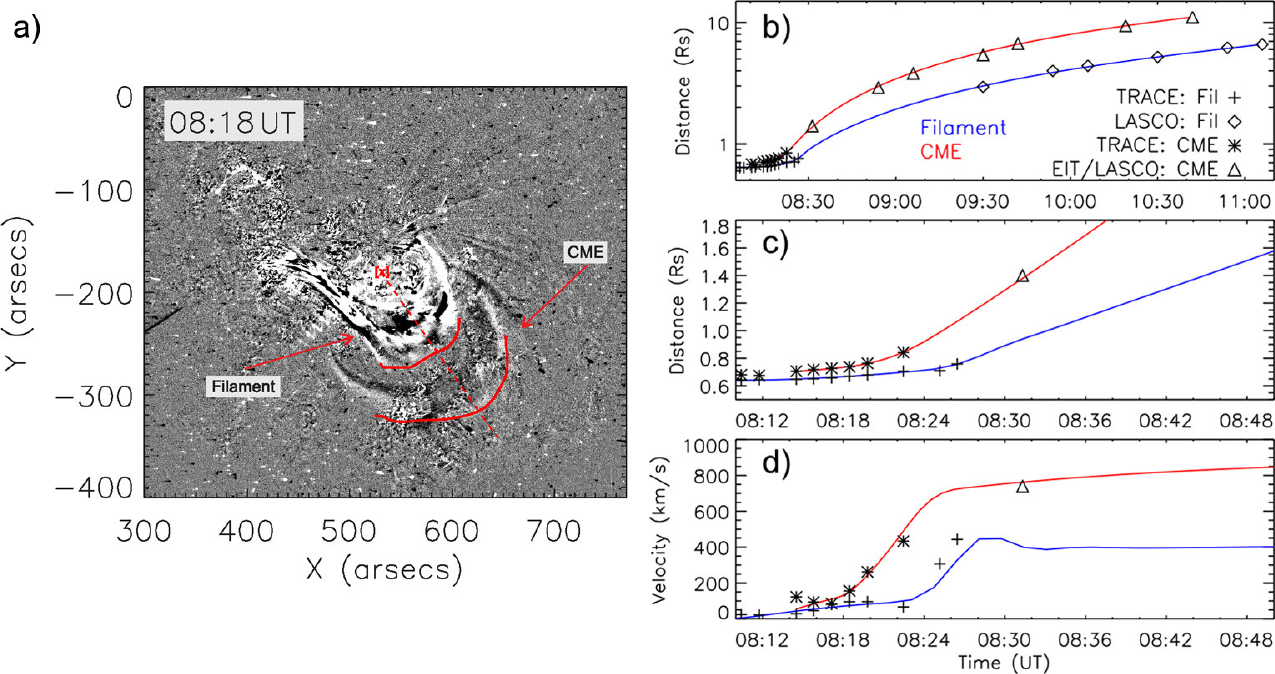}
\caption
{ 
(a) TRACE 171\,\AA ~running-difference image showing the erupting 
filament and the overlying CME front. Distances are measured 
at crossings of the respective leading edges (red solid lines) with the 
dashed line, starting from the point marked by [x]. 
(b) Distance--time plot showing the kinematics of the filament and the
CME front for the entire distance range. 
(c) Distance--time plot for the distance range up to 1.8 $R_\odot$. 
The distance between [x] and the disk centre is added to all TRACE 
and EIT data points. 
(d) Velocity--time plot over the distance range up to 1.8 $R_\odot$. 
See text for further details.
}
\label{fig:obs_traj}
\end{figure}

Figure~\ref{fig:obs_traj}(a) shows a running-difference image from TRACE 171\,\AA ~in which the erupting filament (the CME core) and the preceding CME front can be identified. From a time sequence of similar images by TRACE, EIT, and LASCO C2/C3 we estimated the kinematics of the filament and the CME front, which are shown in Figure~\ref{fig:obs_traj}(b),(c). The distances were measured in the plane of the sky, from disk centre in the LASCO images and from the midpoint of the line connecting the pre-eruption filament footpoints in the TRACE and EIT images. In order to approximately compensate for this discrepancy, we added to the TRACE and EIT measurements the distance between this point and the disk centre, which corresponds to $\approx 400$ Mm. The resulting distances are plotted in Figure~\ref{fig:obs_traj}(b),(c), together with spline--smoothed curves. We did not correct for foreshortening effects as projection effects only result in a multiplication factor and do not alter the profile of the derived kinematical curves \citep[see, \eg,][]{vrsnak07a}. Additionally, Figure~\ref{fig:obs_traj}(d) gives the velocity profiles for the filament and the CME front, as derived from the first derivative of the distance-time measurements and spline--smoothed curves. From these plots we obtain that the coronal loops overlying the filament started their slow rising phase at 08:15 UT, \ie about five\,--\,ten minutes before the filament. Similarly, the CME front reached its final, almost constant velocity a few minutes before the filament.

Various other aspects of the event (flare, CME, EIT wave, dimming) were studied by \cite{jiang07}, \cite{mcintosh07}, \cite{attrill08}, \cite{temmer08}, \cite{miklenic09}, \cite{veronig10}, and \cite{guo.j10}. We refer to these works for further details on the eruption. \cite{guo.j10} suggested that the eruption was triggered by recurrent chromospheric mass injection in the form of surges or jets into the filament channel. Here we propose a different mechanism, assuming that the filament was suspended in the corona by a magnetic flux-rope, a picture that is supported by various magnetic-field models of active regions containing filaments \citep[\eg][]{lionello02,vanballegooijen04,guo.y10a,canou10}. We suggest that the continuous rotation of the sunspot led to a slow expansion of the arcade-like magnetic field overlying the filament (\ie to a continuous weakening of its stabilizing tension), until a critical point was reached at which equilibrium could not be maintained and the flux-rope erupted. We note that we do not claim that the eruption was triggered \textit{exclusively} by this mechanism. Filaments are often observed to spiral into the periphery of sunspots \citep[\eg][]{green07}, and also in our case an inspection of the TRACE and H$\alpha$ images during the early phase of the eruption suggests a possible magnetic connection between the western extension of the filament-carrying core field and the sunspot area. Thus, the sunspot rotation may have added stress to this field, thereby possibly contributing to drive it towards eruption. On the other hand, for an injection of twist as suggested by the simulations mentioned above to occur, the core field must be rooted in the centre of the sunspot, not just in its periphery, which is difficult to establish from observations. It appears reasonable to assume that a clear connection between core field and sunspot centre is not always present, and that the stressing of the overlying ambient field by sunspot rotation may be more relevant for the destabilization of the system in such cases. In order to test this scenario, we perform a series of three-dimensional (3D) MHD simulations, which are described in the following sections.

\section{Numerical Simulations}
\label{sec:num}

The purpose of the numerical simulations presented in this article is to show that the rotation of photospheric flux concentrations can trigger the eruption of an initially stable flux-rope that is embedded in their fields. Differently from previous works \citep[\eg][]{amari96,torok03,aulanier05}, the photospheric vortex motions do not directly affect the flux-rope in our simulations, but solely the field surrounding it.

The first simulation (hereafter run~1) involves a relatively simple magnetic configuration, consisting of a flux-rope embedded in a bipolar potential field (see Figure\,\ref{fig:pfss}(c)). 
The initially potential field gets twisted at its photospheric flux concentrations on both sides of the flux-rope in the same manner. This simulation is very idealized with respect to the observations presented in Section\,\ref{sec:obs}, in particular because both the initial magnetic configuration and the imposed driving possess a high degree of symmetry.

We then consider a more complex initial magnetic field (hereafter run~2), which is chosen such that it resembles the magnetic-field structure prior to the eruption described in Section\,\ref{sec:obs} (see Figure\,\ref{fig:pfss}(d)). As in run~1, this configuration contains a flux-rope embedded in a potential field, but the latter is now constructed by a significantly larger number of sub-photospheric sources, in order to mimic the main features of the observed photospheric flux distribution and the extrapolated coronal magnetic field. Differently from run~1, only one flux concentration is twisted in this case, as suggested by the observations. The purpose of run~2 is to verify that the mechanism studied in run~1 also works in a highly asymmetric configuration. We do not attempt here to model the full eruption and evolution of the CME, for reasons that are specified below. 

To construct our magnetic configurations, we employ the coronal flux-rope model of \citet[][hereafter TD]{titov99}. Its main ingredient is a current ring of major radius [$R$] and minor radius [$a$] that is placed such that its symmetry axis is located at a depth [$d$] below a photospheric plane. The outward directed Lorentz self-force (or ``hoop force'') of the ring is balanced by a potential field created by a pair of sub-photospheric point sources $\pm q$ that are placed at the symmetry axis, at distances $\pm L$ from the ring centre. The resulting coronal field consists of an arched and line-tied flux-rope embedded in an arcade-like potential field. In order to create a shear component of the ambient field, TD added a sub-photospheric line current to the system. Since the latter is not required for equilibrium, we do not use it for our configurations \citep[see also][]{roussev03,torok07}. 

Previous simulations \citep[\eg][]{torok05,schrijver08a} and analytical calculations \citep{isenberg07} have shown that the TD flux-rope can be subject to the ideal MHD helical kink and torus instabilities. Therefore, we adjust the model parameters such that the flux-rope twist stays below the typical threshold of the kink instability for the TD flux-rope \citep[see][]{torok04}. To inhibit the occurrence of the torus instability in the initial configurations, we further adjust the locations and magnitude of the potential field sources such that the field drops sufficiently slowly with height above the flux-rope \citep[see][]{kliem06,torok07,fan07,aulanier10}. While this is a relatively easy task for the standard TD configuration used in run~1, an extended parameter search was required for the complex configuration used in run 2, until an appropriate numerical equilibrium to start with could be found.

\subsection{Numerical Setup}
\label{subsec:num_set}

As in our previous simulations of the TD model \citep[\eg][]{torok04,kliem04}, we integrate the $\beta=0$ compressible ideal MHD equations:
\begin{eqnarray}
\partial_t\rho&=&
                 -\nabla\cdot(\rho\,\bf{u})\,,    \label{eq_rho}\\
\rho\,\partial_{t}\bf{u}&=&
      -\rho\,(\,\bf{u}\cdot\nabla\,)\,\bf{u}
      +\bf{j}\,\mbox{$\times$}\bf{B}
      +\nabla\,\cdot\bf{T}\,,
                                                          \label{eq_mot}\\
\partial_{t}\bf{B}&=&
    \nabla\mbox{$\times$}(\,\bf{u}\,\mbox{$\times$}
    \bf{B}\,)\,,                                  \label{eq_ind}
\end{eqnarray}
where $\bf{B}$, $\bf{u}$, and $\rho$ are the magnetic field, velocity, and mass density, respectively. The current density is given by $\bf{j}=\mu_0^{\,-1}\,\nabla\mbox{$\times$}\bf{B}$. {\bf T} denotes the viscous stress tensor, included to improve numerical stability \citep{torok03}. We neglect thermal pressure and gravity, which is justified for the low corona where the Lorentz force dominates. 

The MHD equations are normalized by quantities derived from a characteristic length [$l$] taken here to be the initial apex height of the axis of the TD current ring above the photospheric plane [$l=R-d$], the maximum magnetic-field strength in the domain [$B_{0 {\mbox{max}}}$], and the Alfv\'en velocity [$v_{a0}$]. The Alfv\'en time is given by [$\tau_a=l/v_{a0}$]. We use a cartesian grid of size $[-40,40] \times [-40,40] \times [0,80]$ for run~1 and $[-40,40] \times [-30,30] \times [0,60]$ for run 2, resolved by $247 \times 247 \times 146$ and $307 \times 257 \times 156$ grid points, respectively. The grids are nonuniform in all directions, with an almost uniform resolution $\Delta = 0.04$ (run~1) and $\Delta = 0.05$ (run~2) in the box centre, where the TD flux-rope and the main polarities are located. The plane $z=0$ corresponds to the photosphere. The TD flux-rope is oriented along the $y$ direction in all runs, with its positive polarity footpoint rooted in the half-plane $y<0$. We employ a modified two-step Lax--Wendroff method for the integration, and we additionally stabilize the calculation by artificial smoothing of all integration variables \citep{sato79,torok03}. 

The boundary conditions are implemented in the ghost layers. The top and lateral boundaries are closed, which is justified given the large size of the simulation box. Below the photospheric plane the tangential velocities are imposed as described in Section~\ref{subsec:num_dri}. The vertical velocities are zero there at all times, and the mass density is fixed at its initial values. The latter condition is not consistent with the imposed vortex flows, but is chosen to ensure numerical stability \citep[see][]{torok03}. Since we use the $\beta=0$ approximation, and since the evolution is driven quasi-statically at the bottom plane, fixing the density in $z=-\Delta z$ is tolerable. The tangential components of the magnetic field [$B_{x,y}$] are extrapolated from the integration domain, and the normal component [$B_z$] is set such that $\nabla \cdot \bf{B} = 0$ in $z=0$ at all times \citep[see][]{torok03}. Since our code does not ensure $\nabla \cdot \bf{B}=0$ to rounding error in the rest of the domain, we use a diffusive $\nabla \cdot \bf{B}$ cleaner \citep[][]{keppens03}, as well as Powell's source-term method \citep[][]{gombosi94}, to minimize unphysical effects resulting from $\nabla \cdot \bf{B}$ errors.

\subsection{Initial Conditions}
\label{subsec:num_ini}

\begin{figure} 
\centering
\includegraphics[width=1.\linewidth]{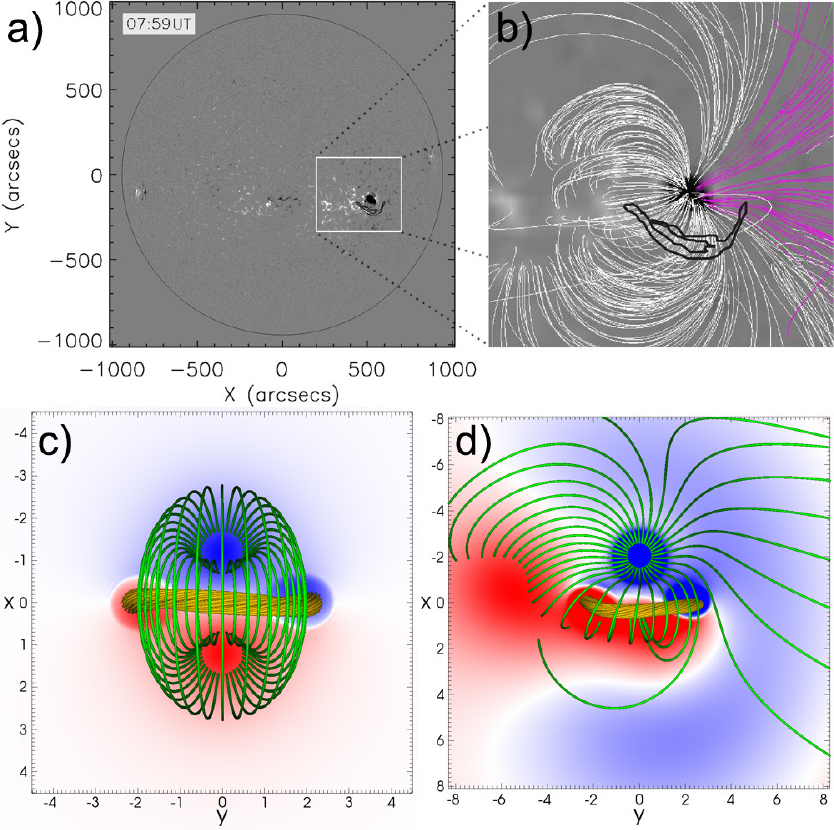}
\caption
{
(a) Same as Figure\,\ref{fig:mdi}. 
(b) Magnetic field lines in the active region area (marked by the white square in panel a) from a Potential 
field source surface (PFSS) model that was calculated for 6 July 2006, 06:04 UT, overlaid on a synoptic 
MDI magnetogram for the corresponding Carrington Rotation 2045. The model corona is a spherical 
shell extending from 1.0 to 2.5 $R_\odot$. Pink (white) field lines depict open (closed) fields. 
The outer contours of the filament, based on H$\alpha$ data taken at 07:59 UT on 6 July 2006, are 
outlined with black lines. For better illustration, the area is rotated to disk centre.
(c),(d): Top view on the magnetic configurations used in runs~1 and 2, respectively, after the initial 
relaxation of the system (see Section\,\ref{sec:num} for details). 
The core of the TD flux-rope is shown by orange field lines, green field lines depict the ambient potential 
field. $B_z$ is shown in the bottom plane, where red (blue) colors corresponds to positive 
(negative) values. The color scale in panel d) is saturated at about 4\,\% of the maximum $B_z$, in order to depict also weaker flux distributions.
}
\label{fig:pfss}
\end{figure}

The parameters of the TD equilibrium employed in run~1 are (in normalized units): $R=2.2$, $a=0.7$, $d=1.2$,  $L=1.2$, and $q=1.27$. The magnetic axis of the TD flux-rope \cite[which is located above the geometrical axis of the current ring, see][]{valori10} has an apex height $z=1.09$. The potential field connects two fully symmetric flux concentrations and runs essentially perpendicular above the TD flux-rope. The apex of the central field line, \ie the field line connecting the centres of the potential-field polarities, is located at $z=3.40$. After the initial relaxation of the system (see below), these heights become $z=1.22$ and $z=3.62$, respectively. Figures\,\ref{fig:pfss}(c) and \ref{fig:rope_simple}(a) show the configuration after the relaxation.

The magnetic configuration used in run~2 is a step towards a more realistic modeling of the coronal field during the 6 July 2006 eruption. Figure~\ref{fig:pfss}(b) shows a coronal potential-field source-surface (PFSS) model \citep{schatten69}, obtained from a synoptic MDI magnetogram for Carrington Rotation 2045, using the {\sf SolarSoft} package {\sf pfss} provided by LMSAL ({\url{http://www.lmsal.com/~derosa/pfsspack/}}). It can be seen that the field lines rooted in the main polarity (the sunspot) form a fan-like structure, which partly overlies the filament. We again consider a standard TD flux-rope, with $R=2.75$, $a=0.8$, and $d=1.75$, but now we use an ensemble of ten sub-photospheric sources \citep[five point sources, and five vertically oriented dipoles like the ones used in][]{torok03} for the construction of the ambient field, in order to resemble the main properties of the observed photospheric flux distribution and the corresponding PFSS field. By adjusting the positions and strengths of the sources, we tried to mimic the approximate flux balance between the concentrated leading negative polarity and the dispersed following positive polarity, the ratio of approximately 10:1 between the peak field strengths in the leading polarity and the following polarity, the size ratio between these polarities, the presence of an ``inverse C-shaped'' area of dispersed negative flux to the West of the leading polarity (see Section\,\ref{sec:obs}), as well as the fan-like shape of the coronal field rooted in the leading polarity. The position of the flux-rope within the ambient field is guided by the observed location of the filament (Figure\,\ref{fig:pfss}(b). Since the model is still relatively idealized, all these features can be matched only approximately. The resulting configuration (after initial relaxation) is shown in Figure\,\ref{fig:pfss}(d) and in Figure\,\ref{fig:rope_real}(a) below. It can be seen that the TD flux-rope is stabilized by flux rooted towards the southern edge of the main polarity. The rope is inclined with respect to the vertical, which is due to the asymmetry of the potential field surrounding it.

In contrast to the configuration used in run~1, the magnetic field in run~2 is dominated by one main polarity. Rather than closing down to an equally strong polarity of opposite sign, the flux emanating from the main polarity now spreads out in all directions, resembling a so-called fan-spine configuration \citep[\eg][]{pariat09,masson09,torok09}. Note that this flux does not contain fully open field lines, as it was presumably the case during the 6 July 2006 eruption (see Figure\,\ref{fig:pfss}(b)). This is due to the fact that the flux distribution shown in Figure\,\ref{fig:pfss}(d) is fully surrounded by weak positive flux in the model (imposed to mimic the isolated ``inverse C-shaped'' weak negative polarity to the West of the main polarity), so that the positive flux in the total simulation domain exceeds the negative flux shown in Figure\,\ref{fig:pfss}(b). Note that this ``total'' flux ratio shall not be confused with the flux ratio between the main polarity and the dispersed positive polarity to its East, which is approximately balanced in the model, in line with the observations.
 
As in \cite{amari96}, \cite{torok03} and \cite{aulanier05}, we use an initial density distribution $\rho_0(\bf{x})=|\bf{B}_0\,(\bf{x})|^{2}$, corresponding to a uniform initial Alfv\'en velocity. For the configuration used in run~2 we also ran a calculation with $\rho_0(\bf{x})=|\bf{B}_0\,(\bf{x})|^{3/2}$, \ie with a more realistic Alfv\'en velocity that decreases with distance from the flux concentrations. We found that the evolution was qualitatively equivalent, but somewhat less dynamic.

In order to obtain a numerical equilibrium as a starting point of the twisting phase, we 
first performed a numerical relaxation of the two configurations used. This is done for
$54\,\tau_a$  for the system used in run~1, and for $75\,\tau_a$ for the system 
used in run~2, after which the time is reset to zero in both cases.

\subsection{Photospheric Driving}
\label{subsec:num_dri}

The velocity field used to twist the potential fields is prescribed in the plane 
$z=-\Delta z$ and located at their main flux concentrations. It produces 
a horizontal counterclockwise rotation, chosen such that the velocity vectors 
always point along the contours of $B_z(x,y,0,t=0)$, which assures that the 
distribution of $B_z(x,y,0,t)$ is conserved to a very good approximation.  
The flows are given by
%
\begin{eqnarray}
u_{x,y}(x,y,-\Delta z,t)&=&v_0f(t)\nabla^\perp\,\{\zeta[B_{0z}(x,y,0,0)]\}\,,
                                                          \label{eq_uxy} \\
u_{z}(x,y,-\Delta z,t)  &=& 0\,.                          \label{eq_uz}
\end{eqnarray}
with $\nabla^\perp:=(\partial_y,-\partial_x)$. A smooth function
%
\begin{eqnarray}
\zeta=B_{z}^2\exp((B_{z}^2-B_{z_{\rm max}}^2)/\delta B^2)\,,
\label{eq_zeta}
\end{eqnarray}
%
chosen as by \cite{amari96}, defines the vortex profile. The parameter $\delta B$ determines the vortex width \citep[see Figure\,3 in][]{aulanier05}. We use $\delta B = 0.7$ for run~1 and $\delta B = 2$ for run~2. The parameter $v_0$ determines the maximum driving velocity. We choose $v_0=0.005$ for both runs to ensure that the driving is slow compared to the Alfv\'en velocity. The velocities are zero at the polarity centre and decrease towards its edge from their maximum value to zero \citep[see Figure\,2 in][]{torok03}. The twist injected by such motions is nearly uniform close to the polarity centre and decreases monotonically towards its edge \citep[see Figure\,\ref{fig:twist_run2} below and Figure\,9 in][]{torok03}. The polarity centres are located at $(\pm 1.2, 0, 0)$ for the configuration used in run~1 and the vortex flows are applied at both flux concentrations. In run~2, we twist the potential field only in the main negative polarity, the centre of which is located at $(-2,0,0)$. The function $f(t)$ describes the temporal profile of the imposed twisting. The twisting phase starts with a linear ramp $(0 \le t \le t_r)$ from $f(0)=0$ to $f(t_r)=1$, which is then held fixed. If a final relaxation phase is added, $f(t)$ is analogously linearly reduced to zero and held fixed. In all simulations in this article $t_r=10\,\tau_a$.

In contrast to the symmetric configuration used in run~1, where most of the flux emanating from the main polarities arches over the flux-rope, the flux that initially stabilizes the rope in run ~2 is concentrated towards the southern edge of the polarity, where the imposed vortex velocities are relatively small. In order to obtain the eruption of the TD rope within a reasonable computational time in run~2, we therefore use a $\delta B$ that is larger than in run~1.

\section{Simulation Results}
\label{sec:res}

\subsection{Run~1}
\label{subsec:run1}

\begin{figure}[t] 
\centering
\includegraphics[width=1.0\linewidth]{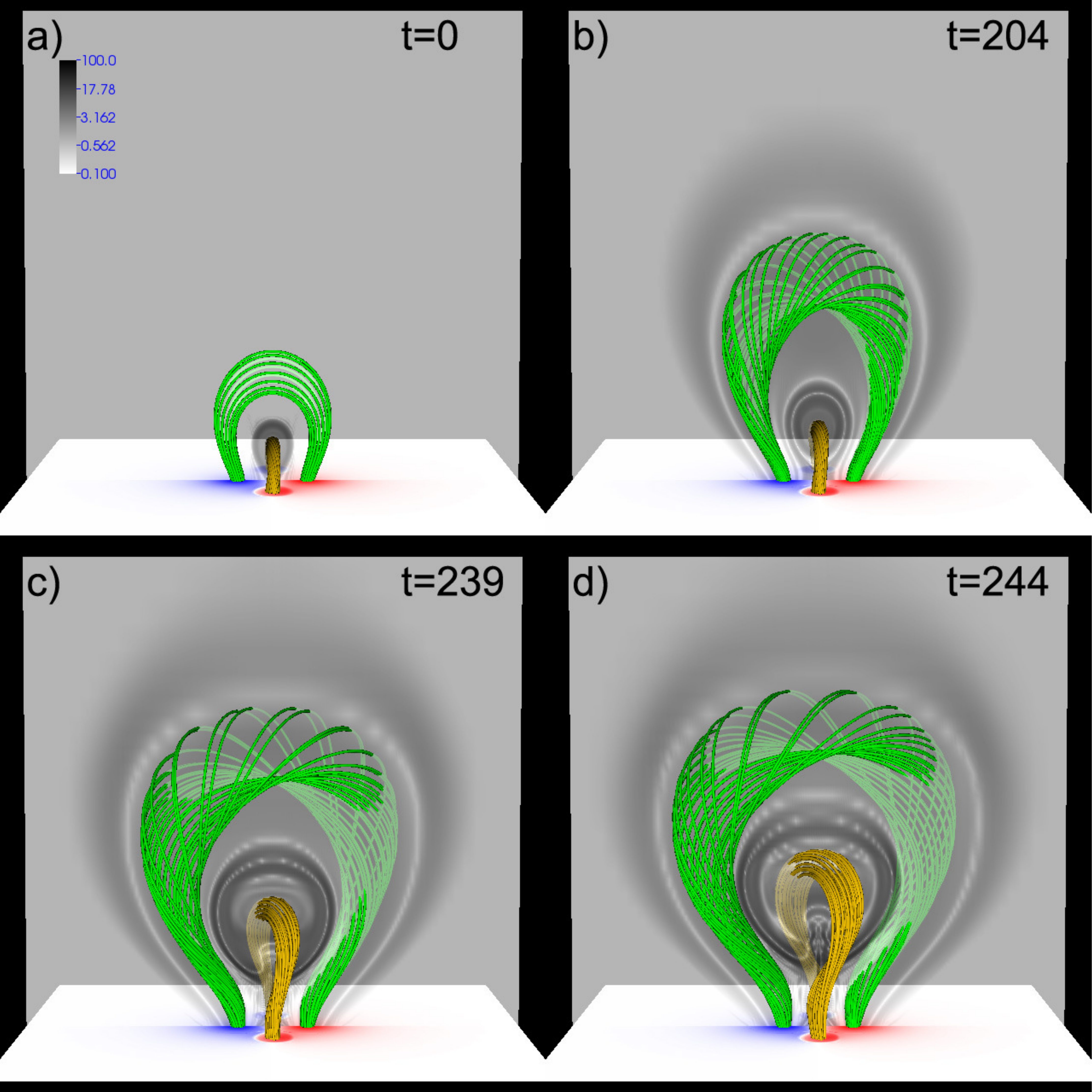}
\caption
{
Magnetic field lines outlining the evolution of the TD flux-rope (orange) and the twisted 
overlying field (green) for run 1, at $t=0, 204, 239, 244\,\tau_A$, respectively; 
panel (a) shows the system after the initial numerical relaxation. The normal component 
of the magnetic field [$B_z$] is shown at the bottom plane $z=0$, with red (blue) 
corresponding to positive (negative) values. The transparent grey-scale shows the 
logarithmic distribution of the current densities divided by the magnetic field strength 
($|\bf{J}| / |\bf{B}|$) in the plane $x=0$. The sub-volume 
$[-8.5,8.5]\times[-8,8]\times[0,16]$ is shown in all panels. An animation of this figure
is available in the electronic version of this article.
} 
\label{fig:rope_simple}
\end{figure}

We first consider the more idealized and symmetric case, in which the vortices are applied at both photospheric polarities of the potential field. As a result of the imposed motions, the field lines rooted in the polarities become increasingly twisted and a relatively wide twisted flux tube is formed, which expands and rises with increasing velocity (Figure\,\ref{fig:rope_simple}). 

\begin{figure} 
\centering
\includegraphics[width=1.0\linewidth]{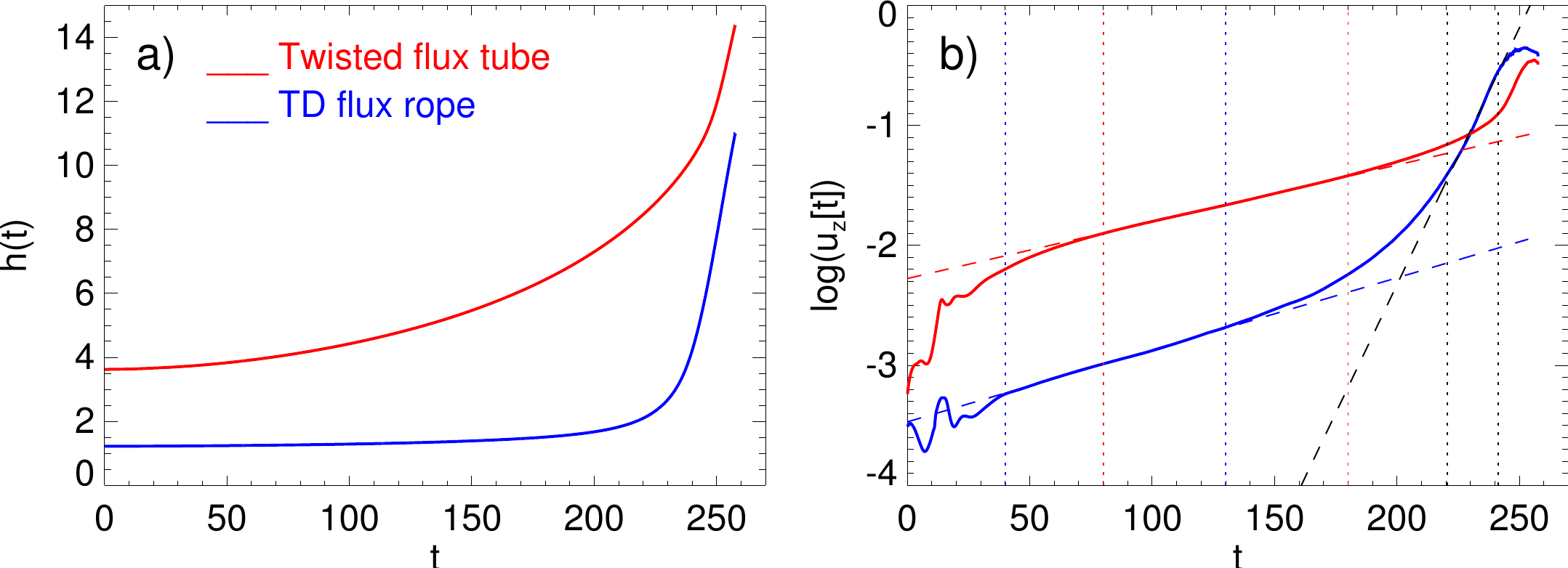}
\caption
{
Kinematics of the TD flux-rope (thick blue lines) and of the overlying 
twisted flux tube (solid red lines) during the twisting phase in run\,1. 
(a) Height of the axis apex as a function of time. The initial heights 
are 1.22 for the TD rope and 3.62 for the twisted flux tube.
(b) Logarithmic presentation of the corresponding vertical velocities. 
The dashed lines show linear fits, obtained within the time periods 
marked by the vertical dotted lines of the same color.
}
\label{fig:cfl}
\end{figure}

Detailed descriptions on the evolution of such twisted fields have been given by \cite{amari96}, \cite{torok03} and \cite{aulanier05}. Since here we are merely interested in how the rising flux affects the stability of the TD flux-rope, we only note that the rise follows the exponential behaviour found in these earlier works. This is shown in Figure\,\ref{fig:cfl}, where the kinematics of the two flux systems (the twisted flux tube and the TD flux-rope) are followed in time by tracking the position of the respective central field line apex. The exponential rise phase of the twisted flux tube, preceded by a slower transition, can be clearly seen between $t \approx 80\,\tau_a$ and $t \approx 180\,\tau_a$.

The slow rise of the flux tube successively weakens the stabilizing magnetic tension on the TD rope, so that the latter starts to ascend as well. As can be seen in Figure\,\ref{fig:cfl}(b), the rise of the TD rope also follows an exponential behaviour up to $t \approx 130\,\tau_a$. While its growth rate is slightly larger than for the twisted flux tube, its velocity remains about one order of magnitude smaller. In order to check that this slow exponential rise of the TD rope is indeed an adaptation to the changing ambient field, rather than a slowly growing instability, we performed a relaxation run by ramping down the photospheric driving velocities to zero between $t=100\,\tau_a$ and $t=110\,\tau_a$ and following the evolution of the system until $t=181\,\tau_a$. Both the twisted flux tube and the TD rope relax towards a numerical equilibrium in this run, without any indication of instability or eruption. Hence, during its slow rise phase until $t \approx 130\,\tau_a$, the TD rope experiences a quasi-static evolution along a sequence of approximately force-free equilibria, generated by the slowly changing boundary conditions (in particular, the changing tangential components of the magnetic field at the bottom plane). 

Starting at $t \approx 130\,\tau_a$, the TD rope undergoes a successively growing acceleration which ends in a rapid exponential acceleration phase between $t \approx 220\,\tau_a$ and $t \approx 250\,\tau_a$ that is characterized by a growth rate significantly larger than during the quasi-static phase (see also the bottom panels of Figure\,\ref{fig:rope_simple}). The rope finally reaches a maximum velocity of $0.45\,v_{a0}$ at $t=252\,\tau_a$, after which it starts to decelerate. Such a slow rise phase, followed by a rapid acceleration, is a well-observed property of many filament eruptions in the early evolution of CMEs \citep[see, \eg,][and references therein]{schrijver08a}, and is also seen for the event studied in this paper (see Figure\,\ref{fig:obs_traj}(d)). The evolution of the TD rope after $t \approx 130\,\tau_a$ can be associated with the development of the torus instability \citep[][]{bateman78,kliem06,demoulin10}, as has been shown under similar conditions in various simulations of erupting flux-ropes \citep{torok07,fan07,schrijver08a,aulanier10,torok11a}.  

During the transition of the TD rope to the torus-unstable regime, the overlying twisted flux tube continues its slow exponential rise at almost the same growth rate for about 100 Alfv\'en times, which excludes the possibility that the additional acceleration of the TD rope after $t \approx 130\,\tau_a$ is due to an adaptation to the evolving environment field. At $t \approx 230\,\tau_a$, however, the rise speed of the TD rope begins to exceed the rise speed of the flux tube, and the latter gets significantly accelerated from below by the strongly expanding rope. The overtaking of the twisted flux tube by the faster TD rope, and the resulting interaction between the two, is reminiscent of the so-called CME cannibalism phenomenon \citep[\eg][]{gopalswamy01,lugaz05}. The investigation of this interaction is, however, beyond the scope of the present paper, so that we stopped the simulation at this point.

Run~1 shows that the rotation of the footpoints of a flux system overlying a stable flux-rope can lead to the eruption of the rope, by progressively lowering the threshold for the torus instability. We suggest that this mechanism may have been at the origin of the CME event described in Section\,\ref{sec:obs}.

The numerical experiment presented here has a high degree of symmetry, with respect to both the initial magnetic field configuration and the driving photospheric motions. A practically identical result is obtained if only one of the polarities of the overlying field is twisted, as long as the driving velocity is clearly sub-alfv\'enic. In particular, we found that twisting only one flux concentration does not significantly affect the rise direction of the TD rope, indicating that slow asymmetric twisting does not necessarily lead to a non-radial rise of the erupting flux-rope if the overlying field is symmetric. A more general case, which exhibits a strongly non-radial rise, is presented in the following section.

\subsection{Run 2}
\label{subsec:run2}

\begin{figure*}[p] 
\centering
\includegraphics[width=1.0\linewidth]{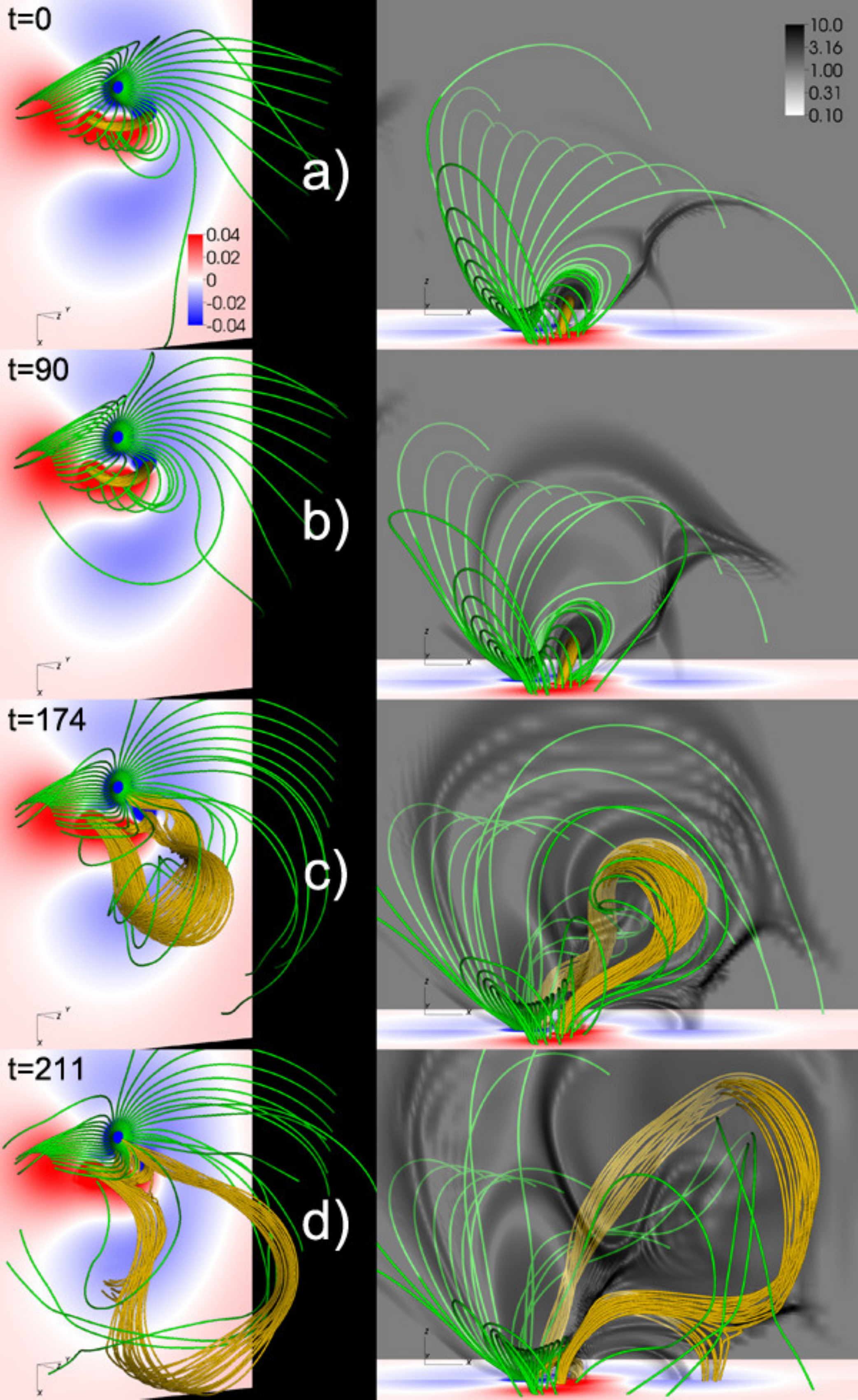}
\caption
{
Snapshots of run~2 at times $t=$ 0, 90, 174, 211\,$\tau_a$, respectively, showing the same features as in Figure\,\ref{fig:pfss}(d). The system is shown after the initial relaxation (a), during the slow rise phase (b), at the time of the peak flux-rope velocity (c), and during the deceleration of the flux-rope (d). The left panels use a view similar to the observations presented in Section\,\ref{sec:obs}, the right panels show a side view. The transparent grey-scale in the right panels depicts the logarithmic distribution of $|\bf{j}|/|\bf{B}|$ in the plane $x=0$, outlining the locations of strong current concentrations. The sub-volume $[-10,16]\times[-11,11]\times[0,18]$ is used for all panels. An animation of this figure is available in the electronic version of this article.
}
\label{fig:rope_real}
\end{figure*}

We now consider a much less symmetric initial condition for the magnetic field, together with a driving that is applied to one polarity only. The configuration is still idealized, but closer to the observations (see Sections\,\ref{sec:obs} and \ref{subsec:num_ini}). The purpose of run~2 is to verify that the CME initiation mechanism suggested in Section\,\ref{subsec:run1} can work also in a more realistic and general setting.

\begin{figure} 
\centering
\includegraphics[width=1.0\linewidth]{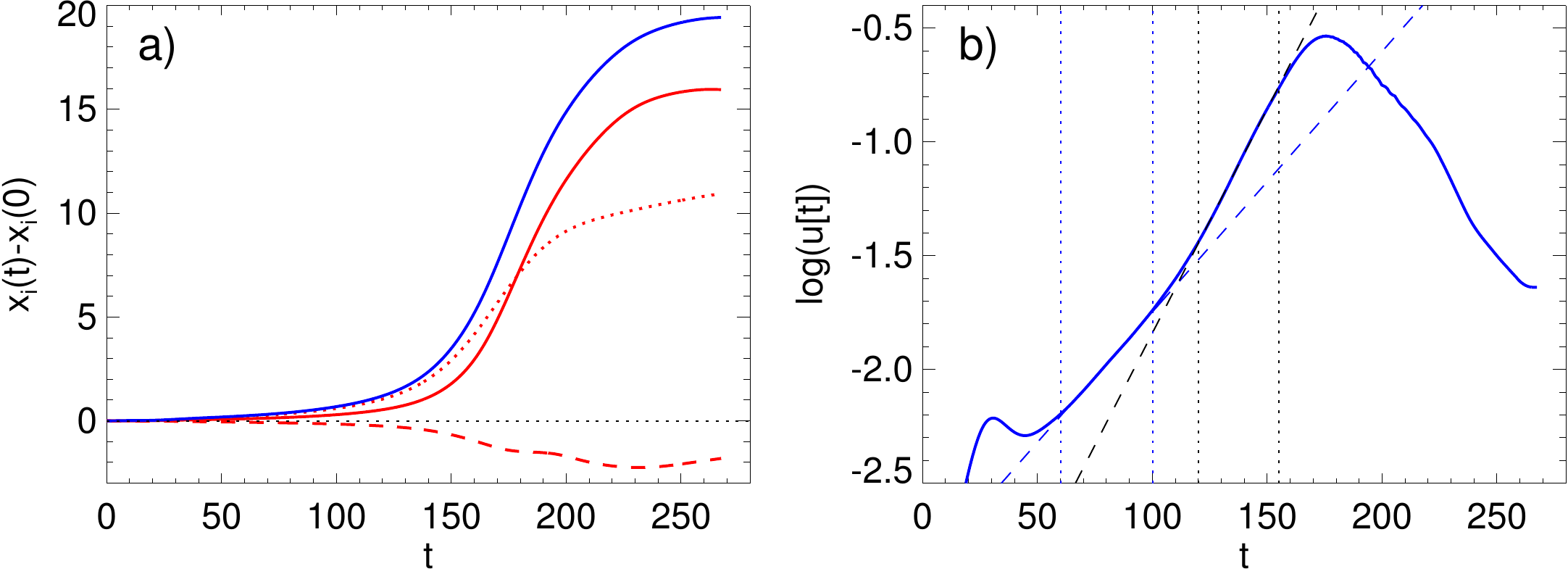}
\caption
{
Kinematics of the TD flux-rope in run~2. 
(a) Red lines show the distances of the axis apex from its initial position, 
$(x_0, y_0, z_0)=(0.075, 0, 1.076)$, for all three spatial dimensions as 
a function of time. The solid line shows $x-x_0$, the dashed one $y-y_0$, 
and the dotted one $z-z_0$. The thick blue line shows the total deviation 
from the initial position.
(b) Logarithmic presentation of the total velocity of the axis apex as a 
function of time (solid blue line). As in Figure\,\ref{fig:cfl}, the dashed 
lines show linear fits obtained for the time periods marked by dotted lines 
of the same color.
}
\label{fig:cfl_run2}
\end{figure}

The fan-like structure of the ambient field makes it difficult to follow its evolution during the twisting phase using a single point as a tracer of the whole three-dimensional structure, as it was done for run~1. We therefore follow here only the apex of the TD rope axis in time. The inclination of the rope makes it complicated to find the exact position of the axis apex, so we determined it only approximately. Consequently, the trajectories presented in Figure\,\ref{fig:cfl_run2} below are somewhat less precise than for run~1.  

Figure\,\ref{fig:rope_real}(a) shows that electric currents are formed in the ambient field volume during the initial relaxation of the system. The strongest current concentrations are located in the front of the flux-rope and exhibit an X-shaped pattern in the vertical cut shown. This pattern outlines the locations of quasi-separatrix layers \citep[QSLs; {\em e.g.}][]{priest92,demoulin96} that separate different flux systems. The QSLs are present in the configuration from the very beginning and arise from the complexity of the potential field (see Section\,\ref{subsec:num_ini}). Their presence is evident also in the left panel of Figure\,\ref{fig:rope_real}a: the green field lines show strong connectivity gradients in the northern part of the main polarity and in the vicinity of the western flux-rope footpoint. It has been demonstrated that current concentrations form preferably at the locations of QSLs and other structural features like null points, separatrix surfaces and separators, if a system containing such structures is dynamically perturbed \citep[\eg][]{baum80,lau90,aulanier05a}. In our case the perturbation results from the -- relatively modest -- dynamics during the initial relaxation of the system.

After the relaxation, at $t=0$, we start twisting the main negative polarity. Due to the pronounced fan-structure of the field rooted in the main polarity, the photospheric twisting does not lead to the formation of a single twisted flux tube that rises exactly in vertical direction above the TD rope, as it was the case in run~1. Rather, the twisting leads to a slow, global expansion of the fan-shaped field lines (see Figure\,\ref{fig:rope_real} and the corresponding online animations). Since we are mainly interested in the destabilization of the flux-rope, we did not study the detailed evolution of the large-scale field. We expect it to be very similar to the one described in \cite{santos11}, since the active region those authors simulated was also dominated by one main polarity (sunspot), and the field rooted therein had a very similar fan-shaped structure (compare, for example, our Figure\,\ref{fig:rope_real} with their Figure\,1).   

Important for our purpose is the evolution of the arcade-like part of the initial potential field that directly overlies the TD flux-rope. Those field lines are directly affected only by a fraction of the boundary flows and therefore get merely sheared (rather than twisted), which still leads to their slow expansion. As it was the case for run~1, the TD rope starts to expand as well, adapting to the successively decreasing magnetic tension of the overlying field. This initial phase of the evolution is depicted in Figure\,\ref{fig:rope_real}b. Note that some of the flux at the front of the expanding arcade reconnects at the QSL current layer (see the online animation), which can be expected to aid the arcade expansion to some degree. As in run~1, the TD rope rises, after some initial adjustment, exponentially during this slow initial phase (Figure\,\ref{fig:cfl_run2}).

As the twisting continues, a transition to a rapid acceleration takes place, which can be seen in Figure\,\ref{fig:cfl_run2}b after $t \approx 100\,\tau_a$, when the rise curve leaves the quasi-static regime. After the transition phase, the TD rope again rises exponentially, but now with a significantly larger growth rate than during the slow rise phase. As for run~1, we attribute this transition and rapid acceleration to the occurrence of the torus instability.

The right panels in Figure\,\ref{fig:rope_real} show that the trajectory of the flux-rope is far from being vertical. As can be seen in Figure\,\ref{fig:cfl_run2}, the rope axis has reached an inclination of about 45 degrees at the time of its peak rise velocity. Such lateral eruptions have been reported frequently in both observations and simulations \citep[][and references therein]{wil05,aulanier10,bi11,panasenco11,zuccarello12,yan.j12}, and are usually attributed to an asymmetric structure of the field overlying the erupting core flux. We believe that this causes the lateral rise also in our case, in particular since we found that asymmetric twisting of a symmetric configuration does not lead to a non-vertical trajectory of the flux-rope (see Section\,\ref{subsec:run1}). We note that such a lateral rise during the very early phase of a CME is different from the often observed deflection of CMEs at coronal holes, where the ejecta is channeled by the structure of the coronal field at larger heights \citep[][and references therein]{kahler12}. As the eruption continues, the trajectory of the flux-rope becomes increasingly horizontal, resembling the so-called ``roll effect'' \citep[\eg][]{panasenco11} and indicating that the rope cannot overcome the tension of the large-scale overlying field. Moreover, as a consequence of its increasing expansion, the flux-rope strongly pushes against the QSL current layer, which results in reconnection between the front of the rope and the ambient field. Eventually, the rope splits into two parts, similar to what has been found in simulations of confined eruptions \citep[][]{amari99a,torok05}. These two effects -- which both are not present in run~1 -- slow down the rise of the rope after $t \approx  175\,\tau_a$ and inhibit its full eruption, \ie the development of a CME) in our simulation.

Since QSLs can affect the evolution of an eruption, but are not expected to play a significant role for its initiation, we did not investigate in detail whether or not QSLs were present in the pre-eruption configuration of the 6 July 2006 event. The PFSS extrapolation indicates their presence to the North and the West of the main polarity (see the field-line connectivities in Figure\,\ref{fig:pfss}(b), but less clearly so to its South. The possible absence of a QSL in front of the erupting core field in the real event is in line with the ``smooth'' evolution of the observed CME, while in our simulation the coherence of the flux-rope is destroyed before it can evolve into a full eruption. Also, the real large-scale field was probably less confining than our model field: the PFSS extrapolation indicates the presence of open field lines, which are fully absent in our simulation.  Since, as stated earlier, we merely aim to model the initiation of the eruption rather than its full evolution into a CME, we refrained from further improving our model to obtain a configuration without a strong QSL in front of the flux-rope and with more open flux.    

As for run~1, we check how the system evolves when the twisting is stopped before the flux-rope erupts. When the vortex flows are ramped down to zero during $t=(35-45)\,\tau_a$ -- corresponding to an effective twisting time of $35\,\tau_a$ -- no eruptive behaviour is seen in the subsequent evolution for almost 300 $\tau_a$, after which we stopped the calculation. However, the system does not fully relax to a numerical equilibrium as it was the case for the simpler configuration (see Section\,\ref{subsec:run1}). Rather, the flux-rope continues to rise very slowly, with velocities smaller than $10^{-3}\,v_{a0}$. This indicates that the system has entered a meta-stable state, which is possibly supported by continuous slow reconnection at the QSL current layer due to numerical diffusion, so that it can be expected that the rope would finally erupt if the integration were continued sufficiently long. When somewhat more twisting is applied, the system behaves as in the continuously driven configuration, \ie a phase of slow rise is followed by a transition to rapid acceleration and the final eruption of the flux-rope, except that the evolution leading up to the eruption takes the longer the less twist is imposed. For example, for an effective twisting period of $45\,\tau_a$, the rapid acceleration of the rope sets in at $\approx 265\,\tau_a$, significantly later than in the continuously driven system.

\begin{figure} 
\centering
\includegraphics[width=0.6\linewidth]{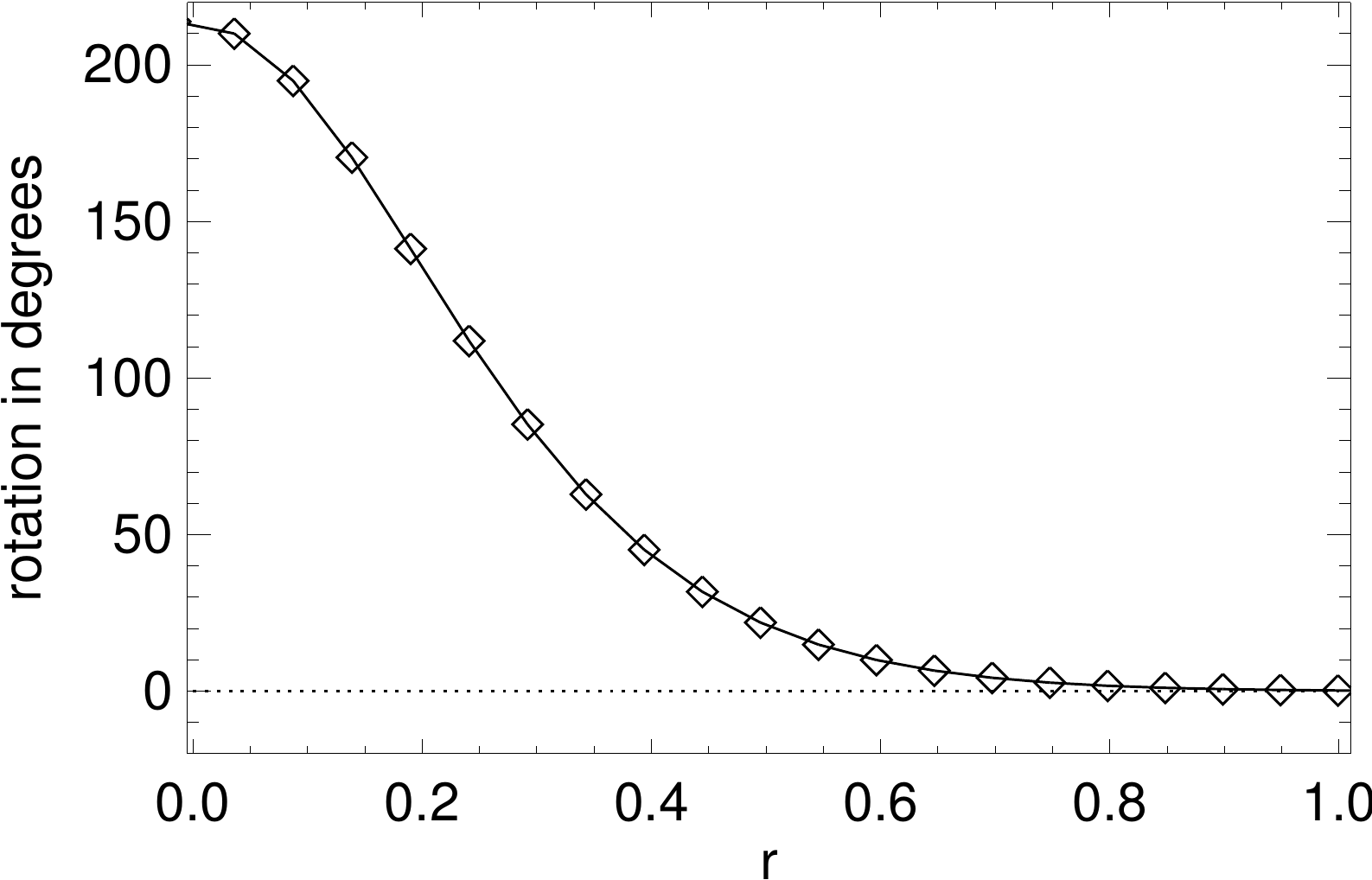}
\caption
{
Rotation profile for run~2 as a function of distance from the vortex centre;
shown at $t=100\,\tau_a$, approximately when the transition from slow to fast 
rise sets in (see Figure\,\ref{fig:cfl_run2}(b)).
}
\label{fig:twist_run2}
\end{figure}

While it is tempting to quantitatively compare the amount of rotation in the simulation with the observed sunspot rotation, we think that such a comparison can be misleading, since the amount of rotation required for eruption will depend on parameters that have not been studied here and are not available from the observations (see the Discussion). Moreover, a quantitative comparison is not straightforward, since the model rotation is highly non-uniform (Figure\,\ref{fig:twist_run2}), while in the observed case a rigid rotation of the spot was measured (Figure\,\ref{fig:spot_evolution}). For example, at $t \approx 100\,\tau_a$, when the transition from slow to fast rise starts in the continuously driven simulation, the field lines rooted very close to the main polarity centre have rotated by about 200$^{\circ}$. However, those field lines do not overlie the TD flux-rope directly, rather they connect to the positive polarity region located to the East of the rope (see Figures\,\ref{fig:pfss}d and \ref{fig:rope_real}) and should therefore not significantly influence the rope's stability. On the other hand, the arcade-like field lines that are located directly above the rope are rooted at a distance of $r \approx 0.4$ from the polarity centre, towards it's southern edge. As can be seen in Figure\,\ref{fig:twist_run2}, the flux surface containing these field lines is rotated by a much smaller amount, about $40^{\circ}$ at $t=100\,\tau_a$. For the run with an effective twisting time of $t=45\,\tau_a$ mentioned above, the imposed total rotation at this flux surface is even smaller, slightly below $20^{\circ}$. These values are similar to the observed sunspot rotation, but, apart from the reasons given above, such a comparison should be taken with care. While the expansion of the field lines located directly above the TD flux-rope presumably depends mainly on the driving imposed at their footpoints, it is also influenced to some degree by the expansion of higher-lying fields which, in turn, depends on the (significantly larger) amount of rotation closer to the polarity centre. Moreover, the values obtained from the model refer to an overlying field that is initially potential (except for the QSL-related current layers), while the real overlying field may have already contained some stress at the onset of detectable rotational motions. Finally, as discussed at the end of the Introduction, the sunspot rotation have have injected stress also directly into the filament. In both cases, presumably less rotation as suggested by the model would have been required to trigger the eruption.

In summary, the simulation successfully models the early phases of the eruption (the slow rise and the initial rapid acceleration of the flux-rope) in a setting that is qualitatively similar to the observed configuration of the active region around the time of the CME described in Section\,\ref{sec:obs}. Hence, the CME-initiation mechanism described in run~1 can work also in more complex and less symmetric configurations.

\section {Summary and Discussion}
\label{sec:dis}

We analyze a filament eruption, two-ribbon flare, and CME that occurred in active region NOAA 10898 on 6 July 2006. The filament was located south of a strong sunspot that dominated the region. In the evolution leading up to the eruption, and for some time after it, a counter-clockwise rotation of the sunspot of about 30$^\circ$ was observed. Similar events, which occurred close to a dominant rotating sunspot, were presented by {\eg} \cite{tian06} and \cite{regnier06}. The triggering of such eruptions is commonly attributed to the injection of twist (or helicity) beyond a certain threshold by the sunspot rotation \citep[\eg][]{torok03}. However, while filaments are frequently observed to spiral into the periphery of main sunspots, the erupting core flux may not always be rooted in the spot itself. Here we suggest that the continuous expansion due to sunspot rotation of the magnetic field that stabilizes the current-carrying core flux, \ie the successive decrease of magnetic tension, can also lead to filament eruptions and CMEs in such configurations.

We support this scenario by MHD simulations, in which a potential field overlying and stabilizing a pre-existing flux-rope is slowly twisted at its photospheric flux concentration(s). The flux-rope is not anchored in these concentrations and is therefore not twisted. In a first configuration, the rope is initially kept in equilibrium by a field rooted in two ``sunspots'' of opposite polarity that are located at opposite sides of the rope. The twisting of the flux concentrations reproduces the known behaviour of twisted bipolar fields (see, \eg, \citealt{amari96}): a twisted flux tube is generated that expands and rises at an exponential rate. As a consequence, the magnetic tension of the field above the pre-existing flux-rope is successively weakened. The rope undergoes a quasi-static adaptation to the changing surrounding field, which manifests in a slow rise phase. As the weakening of the overlying field reaches an appropriate level, the torus instability sets in and rapidly accelerates the rope upwards, leading to a second, fast rise phase and eruption. This evolution in two phases resembles the often observed slow rise phase and subsequent strong acceleration of filaments in the course of their eruption \citep[see Figure\,\ref{fig:obs_traj}, as well as][and references therein]{schrijver08a}. Eventually, since the flux-rope erupts faster than the twisted flux tube rises, the rope catches up and starts to interact with the flux tube, at which point we stop the simulation.

As a step towards more realistic configurations, we consider a second setup in which the initial ambient field surrounding the flux-rope is created by an ensemble of sub-photospheric sources that qualitatively reproduce the photospheric flux distribution and magnetic field structure of the active region around the time of the 6 July 2006 event. In particular, the highly asymmetric flux density and the resulting overall fan shape of the active region field are recovered, while the approximative flux balance of the region is kept. The rotation of the dominant negative polarity (mimicking the observed sunspot rotation) leads to the same qualitative behavior as in the much more symmetric configuration: after a slow rise phase resembling the quasi-static adaptation of the flux-rope to the expanding ambient field, the rope undergoes a second, strong acceleration phase. In this case, the asymmetry of the ambient field leads to a markedly lateral eruption. However, in contrast to the first configuration, the presence of a QSL-related current layer in the front of the erupting flux-rope leads to reconnection which eventually splits the rope before it can evolve into a CME. Although we are not able to follow the expansion of the flux-rope beyond this phase, we can assert the effectiveness of the proposed mechanism in triggering an eruption also in this more realistic case. 

The proposed mechanism requires the presence of a flux-rope in the corona prior to the onset of the twisting motions, which is in line with the relatively small observed rotation of about 30$^\circ$ in our event. Far larger rotations appear to be required to produce a flux-rope that can be driven beyond the threshold of instability by such small additional rotation \citep[\eg][]{torok03,aulanier05,yan.xl12}. It can be expected that the amount of rotation required to initiate the eruption of a pre-existing flux-rope by rotating its overlying field depends on two main parameters: (i) the ``distance'' of the flux-rope from an unstable state and (ii) the ``effectiveness'' of the rotation in reducing the stabilization by the overlying field. For example, it will take a longer time for a low-lying flux-rope to slowly rise to the critical height required for the onset of the torus instability than it does for a rope that is already close to this height. Also, the required rotation will be larger if mostly high-arching field lines, rather than field lines located directly above the rope, are twisted. Thus, the amount of rotation required for eruption appears to depend strongly on the details of the configuration. A proper assessment of this question demands an extensive parametric study that is beyond the scope of this article. Here we merely aim to provide proof-of-concept simulations that illustrate the physical mechanism.

In summary, the main result of our study is that the rotation of sunspots can substantially weaken the magnetic tension of the field in active regions, in particular in cases where the sunspot dominates the region. This can lead to the triggering of eruptions in the vicinity of the spot, even if the erupting core flux (the filament) is not anchored in it. The mechanism that we suggest provides an alternative to the common scenario in which eruptions in the vicinity of rotating sunspots are triggered by the direct injection of twist into the erupting core flux.

\begin{acks}
We thank the anonymous referee for constructive comments that helped to improve the content of the paper. We acknowledge the use of data provided by the SOHO/MDI consortium. SOHO/EIT was funded by CNES, NASA, and the Belgian SPPS. The SOHO/LASCO data used here are produced by a consortium of the Naval Research Laboratory(USA), Max--Planck--Institut f\"ur Aeronomie (Germany), Laboratoire d'Astrophysique de Marseille (France), and the University of Birmingham (UK). SOHO is a mission of international cooperation between ESA and NASA. The {\em Transition Region and Coronal Explorer} (TRACE) is a mission of the Stanford--Lockheed Institute for Space Research, and part of the NASA Small Explorer program. H$\alpha$ data were provided by the Kanzelh\"ohe Observatory, University of Graz, Austria, and by the Hvar Observatory, University of Zagreb, Croatia. The research leading to these results has received funding from the European Commission's Seventh Framework Programme (FP7/2007-2013) under the grant agreements nn$^\circ$ 218816 (SOTERIA project, \url{www.soteria-space.eu}) and n$^\circ$ 284461 (eHEROES, \url{http://soteria-space.eu/eheroes/html}). TT was partially supported by NASA's HTP, LWS, and SR\&T programs. LvDG acknowledges funding through the Hungarian Science Foundation grant OTKA K81421.
\end{acks}


\end{article}
\end{document}